\documentclass[aps,showpacs,secnumarabic]{revtex4}
\usepackage{psfig,epsfig}

\def\ga{\lower.4ex\hbox{$\;\buildrel >\over{\scriptstyle\sim}\;$}}
\def\la{\lower.4ex\hbox{$\;\buildrel <\over{\scriptstyle\sim}\;$}}
\def\half{{\textstyle{1\over2}}}

\def\be{\begin{equation}}
\def\ee{\end{equation}}
\def\bea{\begin{eqnarray}}
\def\nn{\nonumber}
\def\eea{\end{eqnarray}}
\newcommand{\ms}{\noalign{\vspace{3pt plus2pt minus1pt}}}

\def\bi{\bf }
\def\rmF{{\rm F}}

\def\bkm{{|\bf k|}}
\def\bfk{{\bf k}}

\def\pf {p_{\rm F}}
\def\vf {v_{\rm F}}
\def\vef {\varepsilon_{\rm F}}
\def\ve{\varepsilon}

\def\PD1#1{\frac{\partial}{\partial #1}}
\def\pd2#1{\frac{\partial^2}{\partial #1^2}}

\def\calP{\mathcal{P}}
\def\Re{{\rm Re}\, }
\def\Im{{\rm Im}\, }

\begin{document}

\title{Dispersion in a relativistic degenerate electron gas}

\author{J. McOrist$^{1,2}$, D.B. Melrose$^1$  and J.I. Weise$^1$}
\affiliation{$^1$School of Physics, University of Sydney, NSW 2006, AUSTRALIA\\
$^2$Department of Physics, University of Chicago, 5640 S. Ellis Ave., Chicago, IL 60637, USA}

\date{\today}

\begin{abstract}
Relativistic effects on dispersion in a degenerate electron gas are discussed by comparing known response functions derived relativistically (by Jancovici) and nonrelativistically (by Lindhard). The main distinguishing feature is one-photon pair creation, which leads to logarithmic singularities in the response functions. Dispersion curves for longitudinal waves have a similar tongue-like appearance in the relativistic and nonrelativistic case, with the main relativistic effects being on the Fermi speed and the cutoff frequency. For transverse waves the nonrelativistic treatment has a nonphysical feature near the cutoff frequency for large Fermi momenta, and this is attributed to an incorrect treatment of the electron spin. We find (with two important provisos) that  one-photon pair creation is allowed in superdense plasmas, implying relatively strong coupling between transverse waves and pair creation.
\end{abstract}


\maketitle

\section{Introduction}

Dispersion in a completely degenerate electron gas was first studied in connection with plasmons in the context of solid state systems \cite{LLSP,AM76}. Both the longitudinal and transverse response functions for a completely degenerate, nonrelativistic electron gas  were calculated by Lindhard \cite{L54}, and Lindhard's longitudinal response function continues to be used widely. The implied dispersion relation for plasmons is sound-like with the sound speed replaced by the Fermi speed, $\vf$, up to a maximum frequency where it turns over and joins onto the dispersion curve for zero sound \cite{KV91}, giving a tongue-like appearance on a frequency-wavenumber, $\omega$-$|{\bf k}|$, plot. The properties of transverse waves implied by Lindhard's transverse response function seem to have been given relatively little attention. The generalization of the response functions to a completely degenerate, relativistic electron gas was carried out by Jancovici \cite{J62}, and used to discuss relativistic effects  in connection with the longitudinal response \cite{KFH85}, the cutoff frequencies for longitudinal and transverse waves \cite{HM84}, the corrections for a nonzero temperature \cite{MH84}, and fluctuations \cite{S85}. The properties of longitudinal and transverse waves in the interiors of compact stars (white dwarf stars, neutron stars and strange stars) are of relevance to neutrino emission from such dense plasmas; one of the mechanisms for neutrino production, called the plasma process, involves the decay of a wave quantum into a neutrino/anti-neutrino pair \cite{neutrino0}. In this context, in Ref.~\cite{IMHK92} the dispersion relations for these waves were plotted over a restricted range using the Jancovici response functions, and most subsequent authors used the `semi-classical' approximation to the response functions \cite{BS93,neutrinos1,neutrinos2,neutrinos3,neutrinos4}, in which the quantum recoil is neglected. 

In this paper we discuss relativistic quantum effects on dispersion in a completely degenerate electron gas. We concentrate on four specific aspects: (a) the generalization of the properties of longitudinal waves in a nonrelativistic degenerate electrons gas, as described in Ref.~\cite{KV91}, to the relativistic case; (b) the differences between the results derived using the nonrelativistic (Lindhard) and relativistic (Jancovici) forms, particularly for transverse waves; and (c) the validity of the semi-classical approximation, particularly in a superdense plasma, in which the plasma frequency is of order the electron mass (we use natural units, $\hbar=c=1$);  and (d) the possible existence of pair modes.

The response functions can be expressed as integrals over the distribution of particles in momentum space, with the dispersion and dissipation determined primarily by a resonant denominator. The integrals can be expressed such that the denominator is $(\omega-{\bf k}\cdot{\bf v})^2$ in the nonrelativistic, nonquantum case, $(\omega-{\bf k}\cdot{\bf v})^2-|{\bf k}|^4/4m^2$ in the nonrelativistic quantum case, and $(ku)^2-(k^2/2m)^2$ in the relativistic quantum case, with $ku=\gamma(\omega-{\bf k}\cdot{\bf v})$ and $k^2=\omega^2-|{\bf k}|^2$. A characteristic feature of a relativistic quantum treatment is the existence of an additional  source of dissipation and dispersion compared with a nonrelativistic treatment; besides Landau damping (LD), modified by the quantum recoil, one-photon pair creation (PC) is possible. LD is possible only for $\omega<|{\bf k}|$ and PC is possible only for $\omega>(4m^2+|{\bf k}|^2)^{1/2}$, with the intermediate region dissipation free (DF) \cite{T61}. A notable difference between LD and PC is that the contributions from the electron gas to the dissipation have opposite signs: PC occurs in vacuo, and PC due to the electron gas partially (or totally) suppresses the PC (due to  the vacuum) that would otherwise occur. Dispersion is related to dissipation, through causality, and the existence of PC implies the existence of an additional source of dispersion in a relativistic quantum gas. Unlike dissipation, the dispersive effect is not restricted to a particular region of $\omega$-$|{\bf k}|$ space. 

In section~2 we write down the plasma response functions for a degenerate electron gas used in our calculations: the Jancovici, Lindhard and semi-classical forms. In section~3 we plot the response functions, as functions of $\omega$ for fixed $\bkm$, and in section~4 we plot the dispersion relations for longitudinal and transverse waves. We discuss the results in section~5, and our conclusions are summarized in section~6.

\section{Response functions}

We define the response tensor, ${\cal P}^{\mu\nu}(k)$, such that the induced 4-current, $J^\mu(k)$, in the medium is related to the 4-potential, $A^\mu(k)$, by $J^\mu(k)={\cal P}^\mu{}_\nu(k)A^\nu(k)$. In an isotropic plasma, the response may be described in terms of the longitudinal part, ${\cal P}^{L}(k)$, and transverse part, ${\cal P}^{T}(k)$, of the response tensor. In this section we write down these response functions for a completely degenerate, electron gas in the relativistic \cite{J62}, nonrelativistic \cite{L54}, and semi-classical cases.

\subsection{Jancovici's response functions}

The real parts of Jancovici's \cite{J62} response functions in the present notation (with a spurious factor $\omega^2/(\omega^2-|{\bf k}|^2)$ omitted from the transverse response function) are
\bea
{\cal P}^L(k)={e^2\omega^2\over4\pi^2|{\bi k}|^2}\,
\bigg\{{8\varepsilon_{\rm F}p_{\rm F}\over3}
-{2|{\bi k}|^2\over3}\ln\left(
{\varepsilon_{\rm F}+p_{\rm F}
\over m}
\right)
+{\varepsilon_{\rm F}
[4\varepsilon_{\rm F}^2+3(\omega^2-|{\bi k}|^2)]\,
\over6|{\bi k}|}\,
\ln\Lambda_{1\rmF}
\qquad
\nonumber
\\
+{\omega[3|{\bi k}|^2-\omega^2-12\varepsilon_{\rm F}^2]
\over12|{\bi k}|}
\ln\Lambda_{2\rmF}
+{2m^2+\omega^2-|{\bi k}|^2\over3(\omega^2-|{\bi k}|^2)}
|{\bi k}|\varepsilon_k\,
{\omega\over|\omega|}\,
\ln\Lambda_{3\rmF}
\bigg\},
\label{JL}
\eea
\bea
{\cal P}^T(k)=-{e^2\over4\pi^2}\,
\bigg\{{4\omega^2+2|{\bi k}|^2\over3|{\bi k}|^2}\,
\varepsilon_{\rm F}p_{\rm F}
+{2(\omega^2-|{\bi k}|^2)\over3}\ln\left(
{\varepsilon_{\rm F}+p_{\rm F}
\over m}
\right)
\qquad\qquad\qquad\qquad\qquad\qquad\qquad
\nonumber
\\
+\varepsilon_{\rm F}
\left[{\varepsilon_{\rm F}^2(\omega^2-|{\bi k}|^2)\over3|{\bi k}|^3}
+{4m^2|{\bi k}|^2+\omega^4-|{\bi k}|^4\over4|{\bi k}|^3}
\right]
\ln\Lambda_{1\rmF}
\qquad
\nonumber
\\
-{\omega(\omega^2-|{\bi k}|^2)
[12(\varepsilon_{\rm F}^2-\varepsilon_k^2)
+\omega^2+6|{\bi k}|^2]
\over24|{\bi k}|^3}\,
\ln\Lambda_{2\rmF}
-{2m^2+\omega^2-|{\bi k}|^2\over3|{\bi k}|}\,
\varepsilon_k\,
{\omega\over|\omega|}\,
\ln\Lambda_{3\rmF}
\bigg\},
\label{JT}
\eea
where $p_\rmF$ is the Fermi momentum, $\ve_\rmF=(m^2+p_\rmF^2)^{1/2}$ is the Fermi energy,
and with
\be
\varepsilon_k={|{\bf k}|\over2}\left({\omega^2-|{\bf k}|^2-4m^2\over\omega^2-|{\bf k}|^2}
\right)^{1/2}.
\label{vek}
\ee
The factors $\Lambda_{i\rmF}$ are given by setting $|{\bf p}|=p_\rmF$, $\ve=\ve_\rmF$ in
\bea
\Lambda_1={4\varepsilon^2\omega^2-
(\omega^2-|{\bi k}|^2-2|{\bi p}|\,|{\bi k}|)^2
\over
4\varepsilon^2\omega^2-
(\omega^2-|{\bi k}|^2+2|{\bi p}|\,|{\bi k}|)^2},
\qquad
\Lambda_2={4(\varepsilon\omega+|{\bi p}|\,|{\bi k}|)^2
-(\omega^2-|{\bi k}|^2)^2
\over
4(\varepsilon\omega-|{\bi p}|\,|{\bi k}|)^2
-(\omega^2-|{\bi k}|^2)^2},
\nn
\\
\ms
\Lambda_3={(\omega^2-|{\bi k}|^2)^2
(\varepsilon|{\bi k}|+2|{\bi p}|\varepsilon_k)^2
-4m^4\omega^2|{\bi k}|^2
\over
(\omega^2-|{\bi k}|^2)^2
(\varepsilon|{\bi k}|-2|{\bi p}|\varepsilon_k)^2
-4m^4\omega^2|{\bi k}|^2}.
\qquad\qquad
\label{Lambdai3}
\eea

\subsection{Lindhard's response functions}

Lindhard's response functions are \cite{L54}
\bea
{\cal P}^L(k)&=&{e^2n_eu^2\over2mq^2}
\bigg\{1+{1\over2q}
\bigg[1-{1\over4}\left(q-{u\over q}\right)^2\bigg]
\ln\left|{q(q+2)-u\over q(q-2)-u}\right|
\nonumber
\\
&&
\qquad\qquad
+{1\over2q}\bigg[1-{1\over4}\left(q+{u\over q}\right)^2\bigg]
\ln\left|{q(q+2)+u\over q(q-2)+u}\right|
\bigg\},
\qquad\qquad
\label{LL}
\\
{\cal P}^T(k)&=&
-{e^2n_e\over2m}\bigg\{1+{q^2\over4}+{3u^2\over4q^2}-{1\over2q}\bigg[1-{1\over4}\left(q-{u\over q}\right)^2\bigg]^2\ln\left|{q(q+2)-u\over q(q-2)-u}\right|\nonumber\\&&\qquad\qquad-{1\over2q}\bigg[1-{1\over4}\left(q+{u\over q}\right)^2\bigg]^2\ln\left|{q(q+2)+u\over q(q-2)+u}\right|\bigg\},
\qquad\qquad
\label{LT}
\eea
with
$u={2m\omega/p_{\rm F}^2}$, $q={|{\bi k}|/p_{\rm F}}$ and $n_e=\pf^3/3\pi^2$. Lindhard's response functions may be obtained from Jancovici's response functions by expanding in $\bkm/\pf$ and $\omega/m$ for $\pf\ll m$.

\subsection{Semi-classical approximation}

The semi-classical approximation is obtained by neglecting the quantum recoil terms in the resonant denominator. They are
\bea
{\cal P}^L(k)&=&{\omega_c^2\over\mu_0}\,\frac{3\omega^2}{\bkm^2\vf^2}
\left(\frac{\omega}{2\bkm\vf}\ln\frac{\omega+\bkm\vf} {\omega-\bkm\vf}-1\right),
\label{BSL}
\\
{\cal P}^T(k)&=&{\omega_c^2\over\mu_0}\, \frac{3\omega^2}{2\bkm^2\vf^2}
\left(1-\frac {\omega^2-\bkm^2\vf^2}{\omega^2}
\frac{\omega}{2\bkm\vf}\ln\frac{\omega+\bkm\vf} {\omega-\bkm\vf}\right),
\label{BST}
\eea
with $\omega_c$ the cutoff frequency (referred to as the plasma frequency by some authors), and $\vf=\pf/\vef$ the Fermi speed. Approximate forms similar to (\ref{BSL}) and (\ref{BST}) were given by Lindhard (1954) and Jancovici (1962). In Ref.~\cite{BS93} the additional approximation $\vf\to1$ is made in (\ref{BSL}) and (\ref{BST}), corresponding to a highly relativistic electron gas.

\subsection{Response functions for the vacuum}

The  vacuum polarization tensor is of the form 
${\cal P}^{\mu\nu}(k)={\cal P}_0(k)[g^{\mu\nu}-k^\mu k^\nu/k^2]$, and the form of ${\cal P}_0(k)$ is well known \cite{LLRQM}. The longitudinal and transverse parts of the vacuum response tensor are
${\cal P}_0^L(k)=\omega^2{\cal P}_0(k)/(\omega^2-|{\bf k}|^2)$ and ${\cal P}_0^T(k)={\cal P}_0(k)$, respectively, which are to be added to the contributions from the electron gas. The real part is unimportant for present purposes, but it is essential to include the imaginary part of vacuum polarization when discussing dissipation due to PC. It has the specific form \cite{LLRQM}
\be
\Im{\cal P}_0(k)=
{e^2\over3\pi}\,
\big[m^2+\half(\omega^2-|{\bi k}|^2)\big]\,\frac{\ve_k}{\bkm},
\label{Imvac}
\ee
for $\omega^2-|{\bi k}|^2>4m^2$, with $\Im{\cal P}_0(k)=0$ for $\omega^2-|{\bi k}|^2<4m^2$.

\begin{center}

\begin{figure}[hp]
\centerline{\includegraphics[width=\columnwidth]{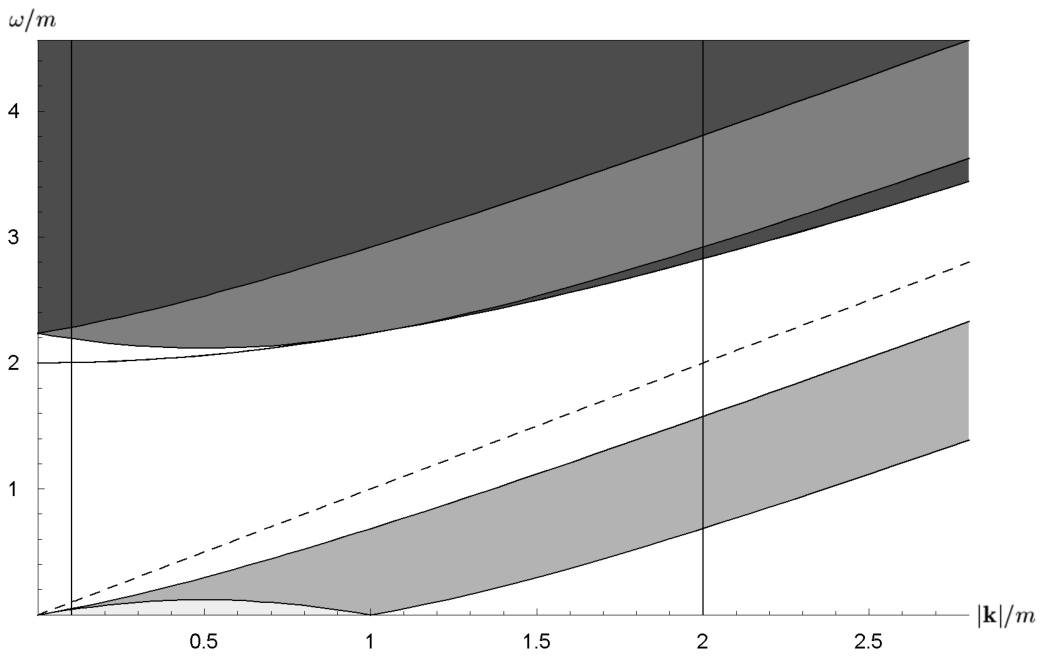}}
\caption{The regions where dissipation is allowed in a degenerate electron gas with $\pf/m = 0.5$ are shown shaded. The unshaded areas represent the DF regions. LD is allowed only to the right of the light line (small dashed) and is further restricted to the shaded area by the Pauli exclusion principle. PC is allowed only to the left of the curve $\omega = (4m^2 + \bkm^2)^{1/2}$, where dissipation in the vacuum is allowed (darkest shading) and this is completely suppressed by the electron gas in the unshaded area indicated. The vertical lines indicate the two values of $|{\bf k}|/m=0.1, 2$ chosen to plot the response function.}
\label{fig:thresholds}
\end{figure}
\end{center}

\subsection{Allowed regions for dissipation}

The allowed regions for LD and PC are illustrated in Figure~\ref{fig:thresholds} for $\pf/m = 0.5$. Of particular relevance to the discussion below is the region where PC is allowed. PC is allowed in principle only for $\omega>(4m^2+\bkm^2)^{1/2}$. The completely degenerate electron gas suppresses PC completely in a small contiguous region at higher $\omega$ and small $\bkm$, corresponding to $(4m^2+\bkm^2)^{1/2}<\omega<\vef+\ve_{\pf-\bkm}$, where $\ve_{\pf\pm\bkm}=[m^2+(\pf\pm\bkm)^2]^{1/2}$ are threshold energies. For $\bkm=0$ this precludes PC for {\bf $\omega<2\vef$}, as pointed out in Ref.\ \cite{IMHK92}. However, the region at $\omega>(4m^2+\bkm^2)^{1/2}$ where PC is completely suppressed by the degenerate electron gas is restricted to $\bkm<2\pf$. A closely analogous situation applies to PC in a magnetized electron gas, and plot similar to Figure~\ref{fig:thresholds} was used in Ref.~\cite{PK92} to illustrate this case.

\begin{figure}[hp]
\includegraphics[width=5in]{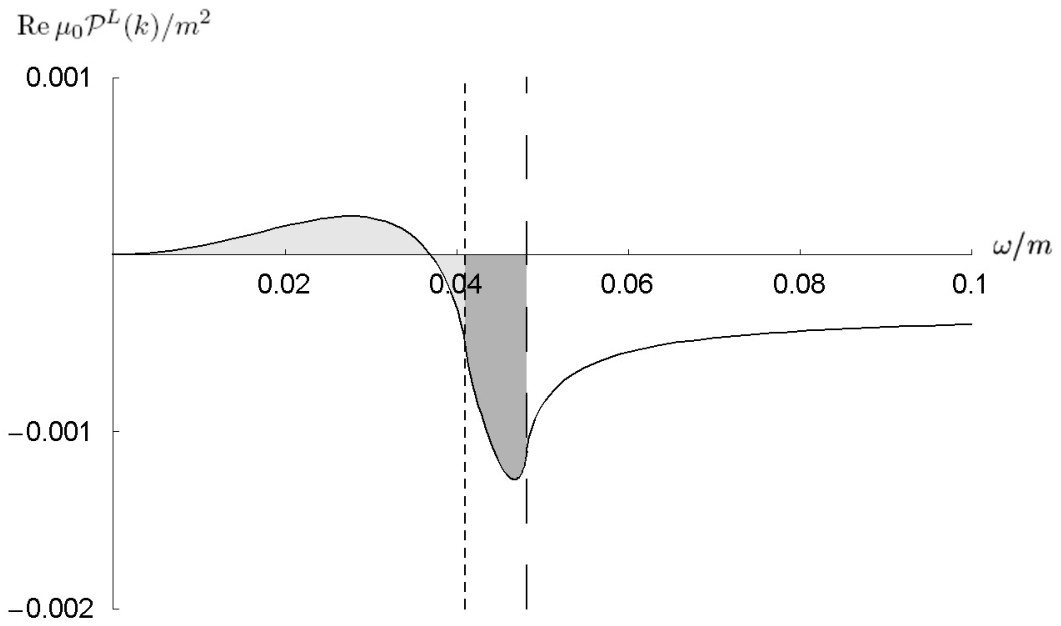}
\includegraphics[width=5in]{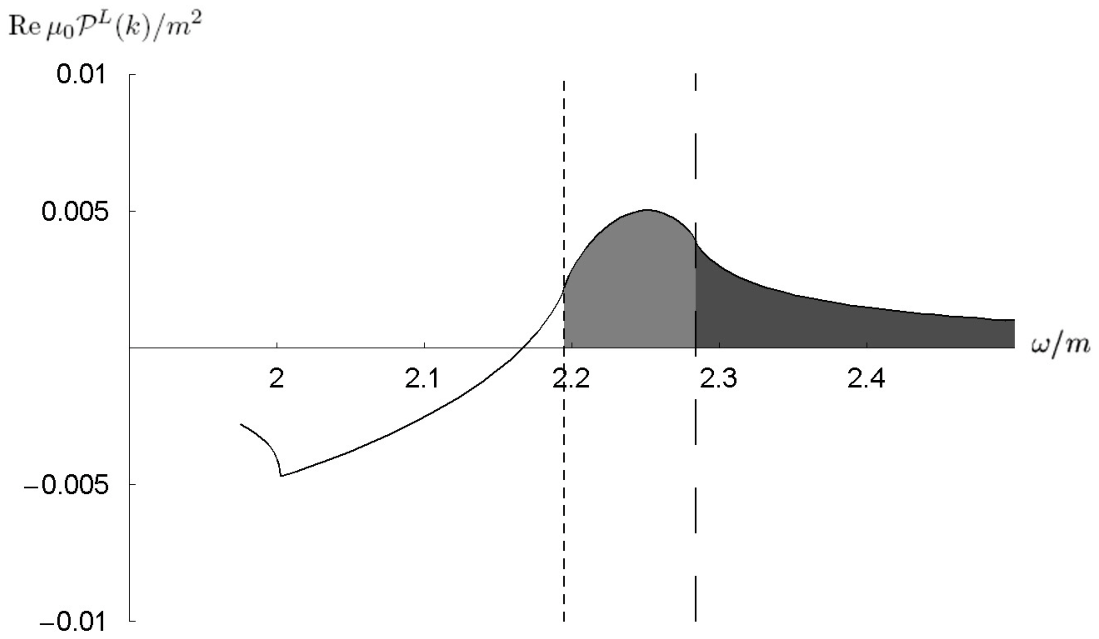}
\caption{$\Re\mu_0 \calP^L(k)$ is shown as a function of $\omega$ for $\bkm= 0.1m$, separated into two parts: (a) the LD regime ($\omega<|{\bf k}|$) where dissipation occurs in the shaded regions and is due to the degenerate electron gas, and (b) the PC regime ($\omega>(4m^2+|{\bf k}|^2)^{1/2}$) where PC due to the degenerate gas completely suppresses the vacuum PC below the short dashed line, partially suppresses the vacuum PC between the two dashed lines, and makes no contribution to the vacuum PC above the long dashed line.}
\label{fig:janco_long_thresholds}
\end{figure}

\section{Plots of the response functions}

Figure~\ref{fig:janco_long_thresholds} shows plots for $\bkm = 0.1m$.  The upper and lower panels show the response function on two different scales, with the upper panel emphasizing the region where LD is allowed and the dispersion is dominated by that associated (by the Kramer-Kronig relations) with LD, and with the lower panel emphasizing the region where PC is allowed and where dispersion associated with PC is important. The regions where LD and PC are allowed are shown shaded in the two panels, respectively. The short-dashed and long-dashed vertical lines correspond to $\omega=|\vef-\ve_{\pf-\bkm}|$ and $\omega=\ve_{\pf-\bkm}-\vef$, respectively, in the upper panel,  and to $\omega=\vef+\ve_{\pf-\bkm}$  and $\omega=\vef+\ve_{\pf+\bkm}$, respectively, in the lower panel.  These values of $\omega$ define the boundaries between the differently shaded regions within the LD and PC regimes in Figure~\ref{fig:thresholds}. In general, for $\bkm<2\pf$, LD is possible for $\omega<\ve_{\pf+\bkm}-\vef$, and it changes its analytic form at $\omega=\vef-\ve_{\pf-\bkm}$. PC due to the electron gas is possible only for $\vef+\ve_{\pf-\bkm}<\omega<\vef+\ve_{\pf+\bkm}$, where it partially or totally suppressed PC due to the vacuum.

\begin{figure}[hp] 
\includegraphics[width=5in]{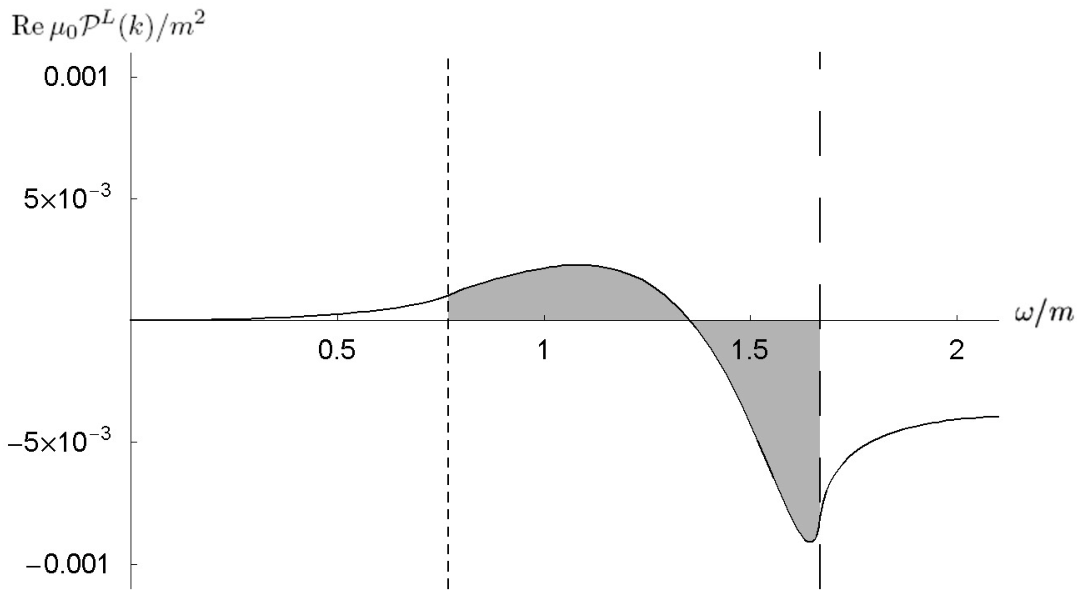}
\includegraphics[width=5in]{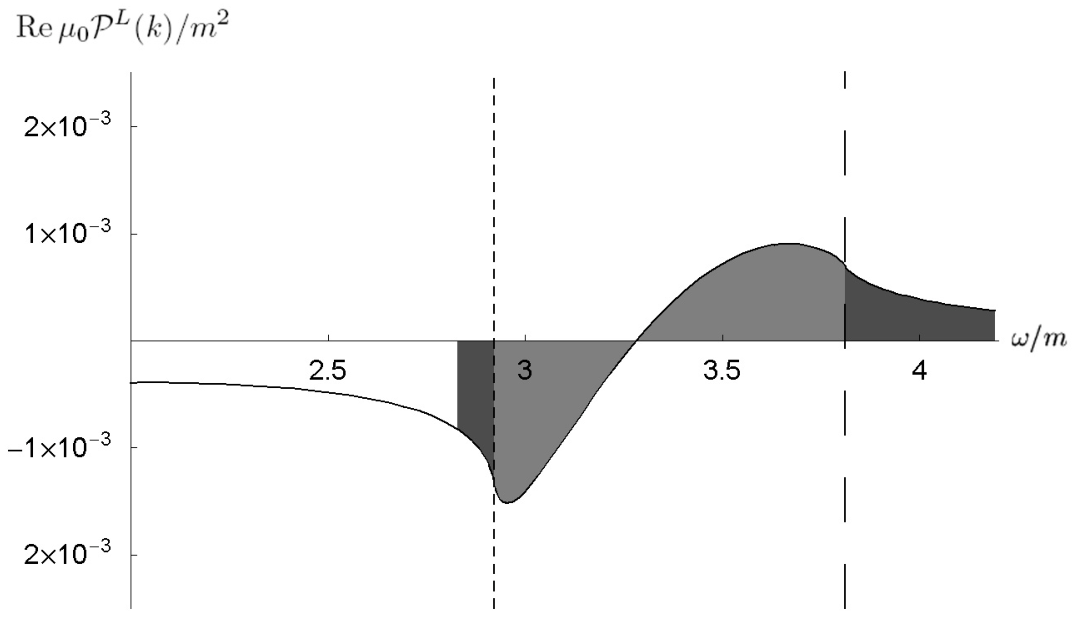}
\caption{$\Re \mu_0\calP^L(k)$ is shown as a function of $\omega$ for $\bkm= 2m$, separated into two parts: (a) the regime $\omega<|{\bf k}|$, with LD is nonzero only between the two dashed lines, and (b) the regime $\omega>(4m^2+|{\bf k}|^2)^{1/2}$, with PC due to the degenerate gas nonzero only between the two dashed lines.}
\label{fig:janco_long_thresholds2}
\end{figure}

Figure~\ref{fig:janco_long_thresholds2} shows plots of $\Re\mu_0\calP^L/m^2$ as a function of $\omega$ for $\bkm = 2m$.  In Figure~\ref{fig:janco_long_thresholds}(a), LD is only allowed between the vertical lines.  In Figure~\ref{fig:janco_long_thresholds}(b), PC is allowed for $\omega>(4m^2+\bkm^2)^{1/2}$.  The vacuum PC is partially suppressed by the contribution of the electrons between the short and long dashed lines.

\begin{figure}[hp]
\includegraphics[width=4.7in]{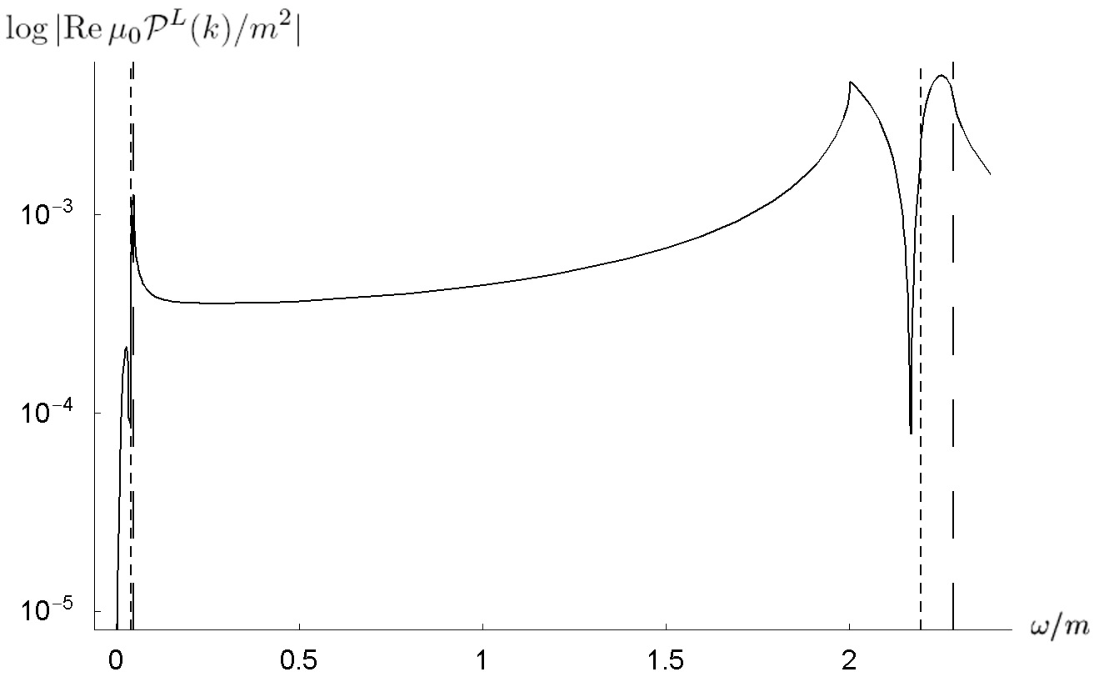}
\includegraphics[width=4.7in]{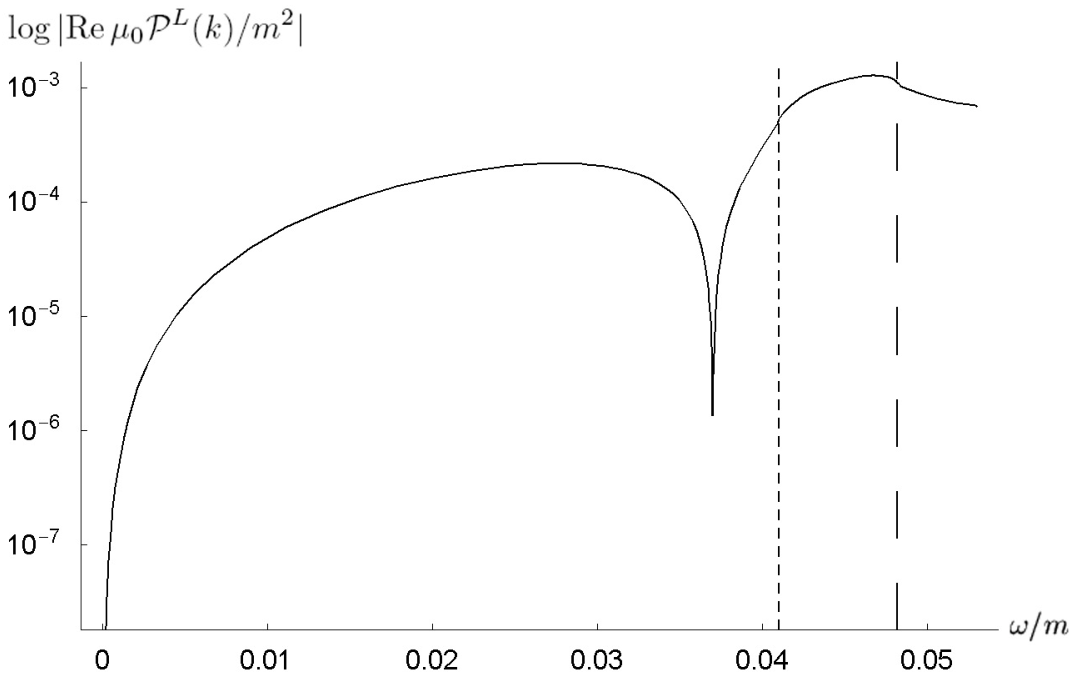}
\includegraphics[width=4.7in]{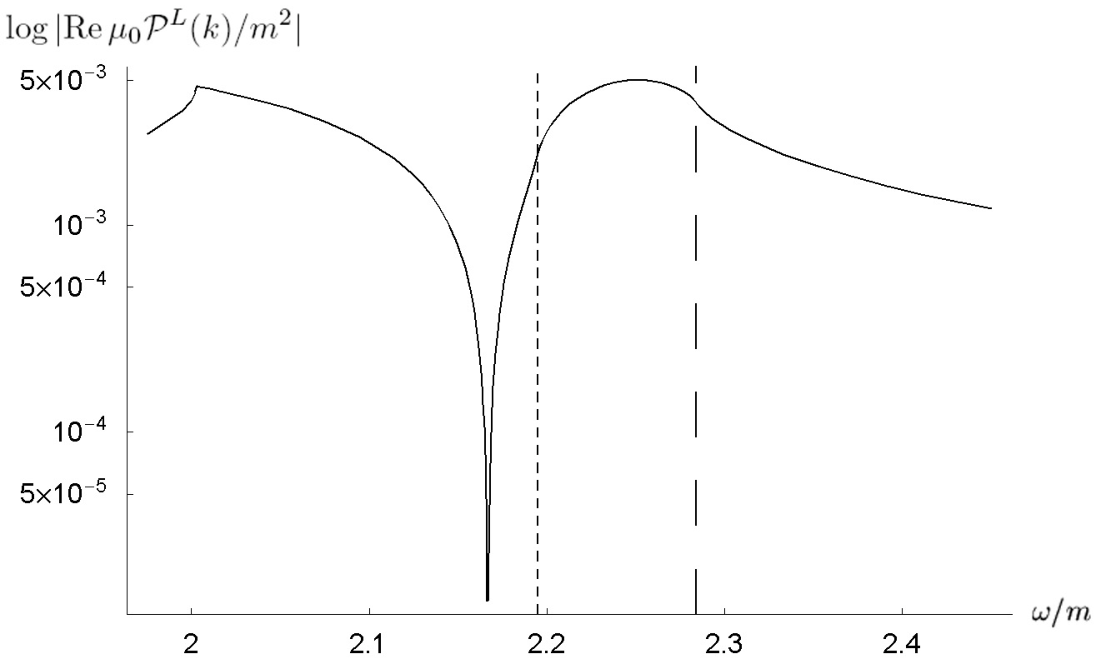}
\caption{A plot of $\log |\Re\mu_0\calP^L(k)/m^2|$ for $\pf/m = 0.5$ and $\bkm/m = 0.1$. (a) The overall form of the response function; the cusps correspond to points where $\Re\mu_0\calP^L(k)$ passes through zero. (b) The LD region in more detail. (c) The PC region in more detail. The vertical lines define the different regions in the LD and PC regimes (see text).}
\label{fig:janco_long_log}
\end{figure}

The response function varies greatly in magnitude over the entire frequency range, and in Figure \ref{fig:janco_long_log} we plot $\log |\Re\mu_0\calP^L(k)/m^2|$ to show the changes on the larger scale. Figure \ref{fig:janco_long_log}(a), (b) and (c), show the logarithm over the entire frequency range and over ranges similar to those in Figures \ref{fig:janco_long_thresholds}(a), (b), respectively. The logarithm introduces singularities at points where $\Re\mu_0\calP^L(k)$ passes through zero, as is evident by noting that the downward-pointing cusp-like feature in Figure \ref{fig:janco_long_log}(b) corresponds to the point where the response function changes sign in Figure \ref{fig:janco_long_thresholds}(a).

\subsection{Transverse response function}

\begin{figure}[hp]
\includegraphics[width=4.7in]{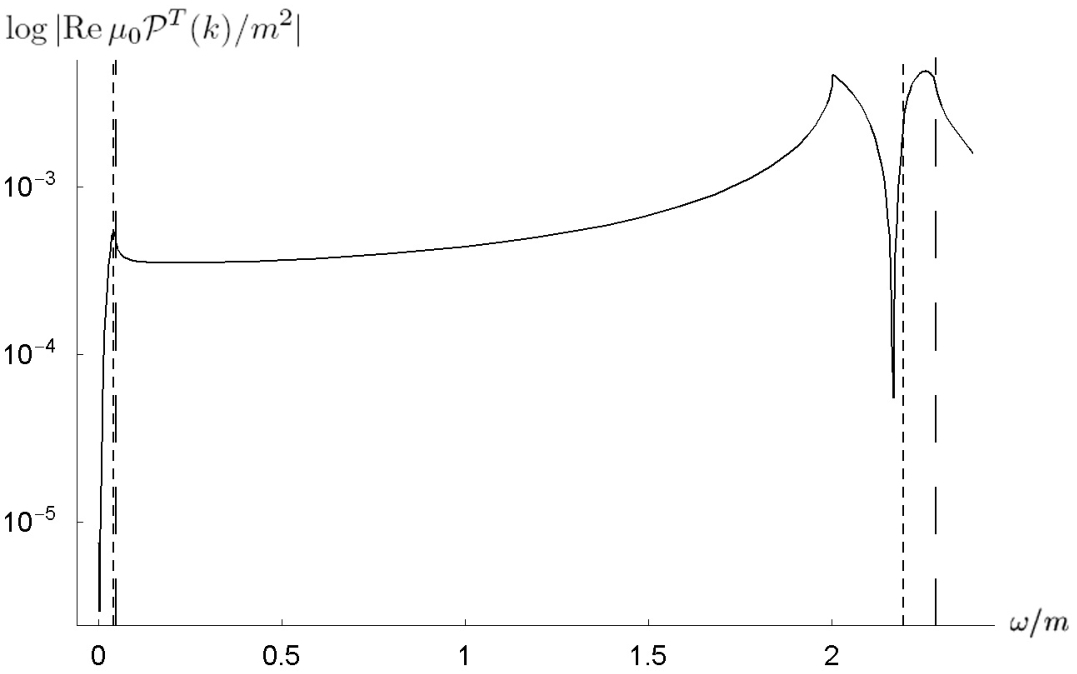}
\includegraphics[width=4.7in]{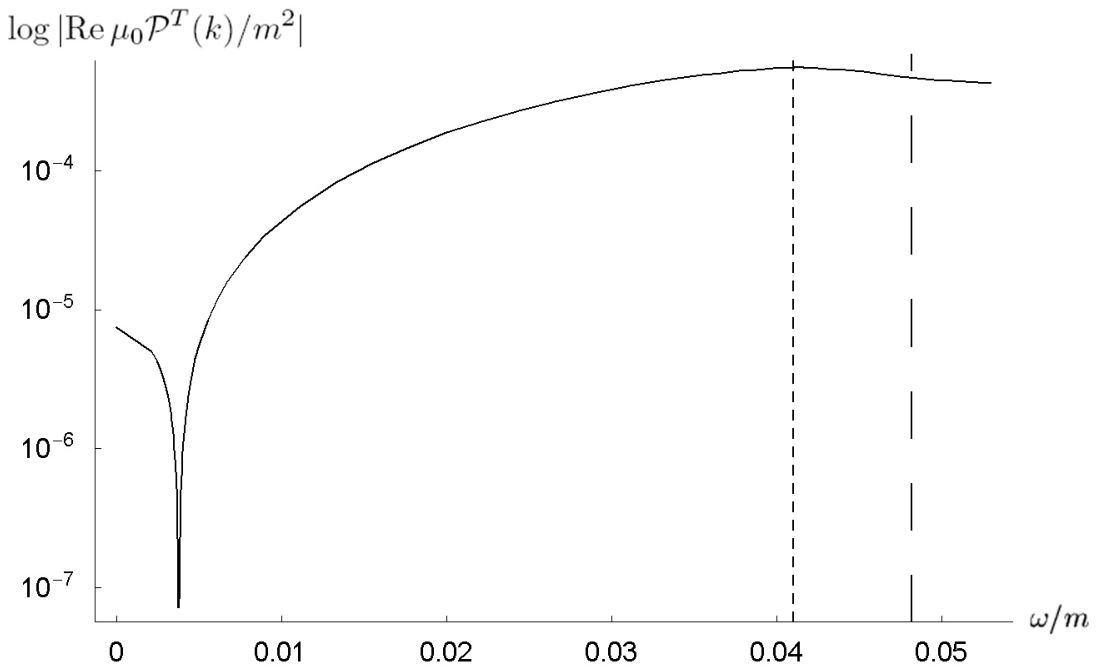}
\includegraphics[width=4.7in]{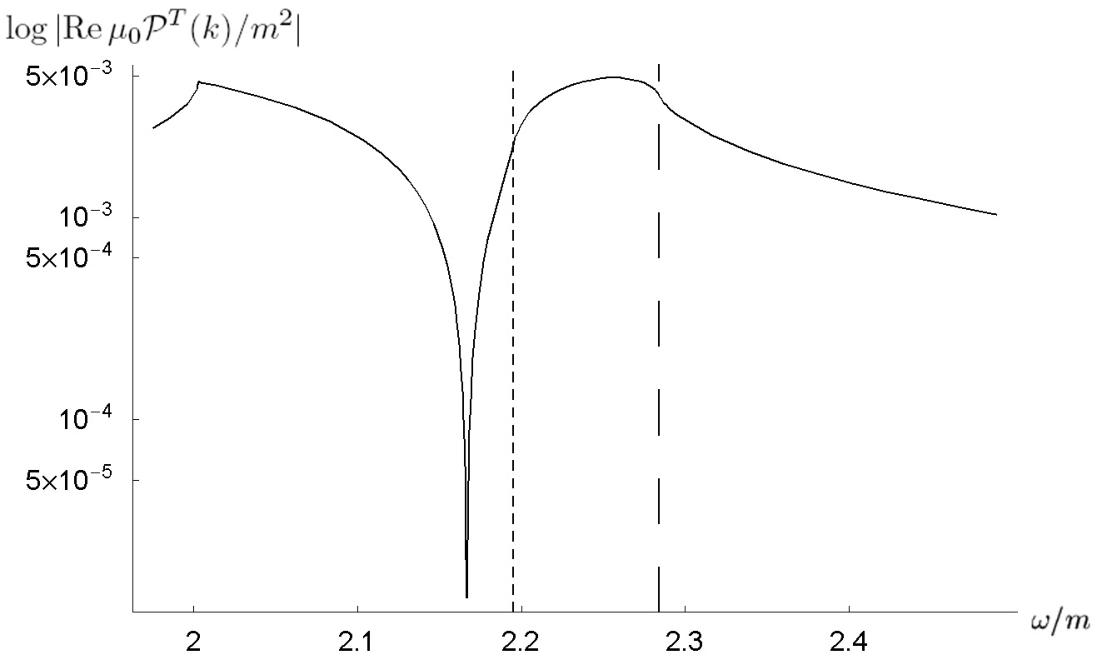}
\caption{As for Figure~4, but for $\log |\mu_0\Re\calP^T(k)/m^2|$.}
\label{fig:janco_trans_log}
\end{figure}

The transverse response function exhibits features similar to the longitudinal response function. In Figure \ref{fig:janco_trans_log} $\log|\Re\mu_0\calP^T(k)/m^2|$ is plotted in the format used in Figure \ref{fig:janco_long_log}. The downward pointing cusp-like feature in Figure \ref{fig:janco_trans_log}(b) corresponds to a change in sign of the response function, from positive at low frequency to negative at high frequency. The vertical lines correspond to the boundaries of the allowed dissipation regions, and are the same as in Figure \ref{fig:janco_long_log}.

\subsection{Comparison of Jancovici, Lindhard and semi-classical forms}

\begin{figure}[hp]
\includegraphics[width=4.7in]{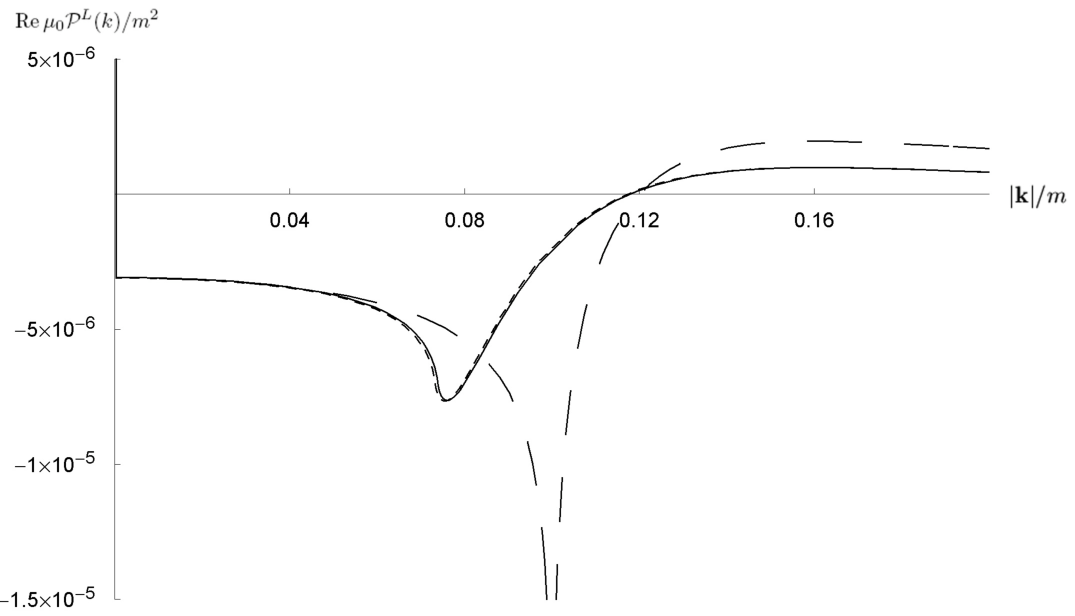}
\includegraphics[width=4.7in]{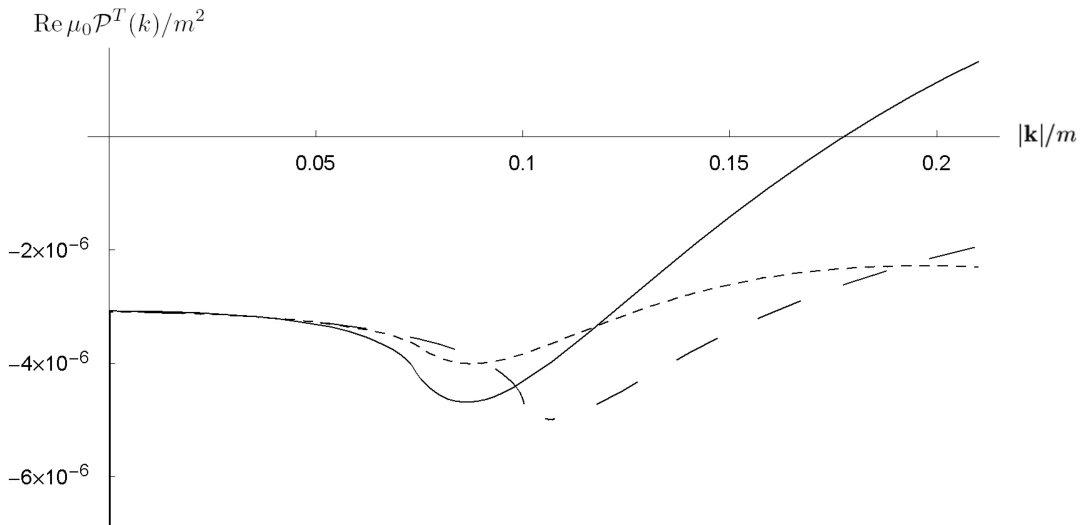}
\caption{The longitudinal and transverse response functions are plotted for $\omega/m=0.01$ and $\pf/m=0.1$, as a function of $\bkm/m$ for the Jancovici (solid curves), Lindhard (dotted curves) and semi-classical (dashed curves) forms.}
\label{fig:allthree}
\end{figure}

In Figure~\ref{fig:allthree} we compare the longitudinal and transverse response functions for the three different forms written down in section~2: the Jancovici form, (\ref{JL}) and (\ref{JT}), the Lindhard  form, (\ref{LL}) and (\ref{LT}), and the semi-classical form, (\ref{BSL}) and (\ref{BST}).  The values chosen are $\omega/m=0.01$ and $\pf/m=0.1$, and the response functions are plotted as a function of $\bkm/m$. For $\bkm<\omega$ all three forms give similar results. The Lindhard form deviates substantially from the Jancovici form for $\bkm\ga5\omega$, and the Lindhard form becomes an increasingly poor approximation as $\bkm/\omega$ increases further. 

The important requirements for the semi-classical approximation to be accurate are $\omega/\vef\ll1$ and $\bkm\ll2\pf$. For the longitudinal response, the semi-classical result is a poor approximation for large $\bkm\ll2\pf$ but the Lindhard form is accurate for arbitrarily large $\bkm/2\pf$, provided one has $\omega\ll m$, $\pf\ll m$. For the transverse response, both the semi-classical and the Lindhard forms are poor approximations for sufficiently large $\bkm/2\pf$.

\section{Wave dispersion}

The dispersion equations for longitudinal and transverse modes in the rest frame of the plasma are
\be
\omega^2+\mu_0{\cal P}^L(k)=0,
\qquad
\omega^2-|{\bf k}|^2+\mu_0{\cal P}^T(k)=0,
\label{deqn1}
\ee
respectively. In this section we give numerical and analytic results for the longitudinal and transverse modes in a completely degenerate electron gas. 

\subsection{Longitudinal modes}
The longitudinal mode of a thermal electron gas is usually called the Langmuir mode. For longitudinal waves in a nonrelativistic  degenerate electron gas, there is a highest frequency \cite{SU82}, where there is a turnover of the dispersion relation \cite{KV91}. In the discussion here we concentrate on the relativistic generalization of this known nonrelativistic result. Specifically, we compare the dispersion relation for Langmuir waves calculated using Lindhard's response function, with that calculated using Jancovici's response function.

\begin{figure}[hp]
\includegraphics[width=4.5in]{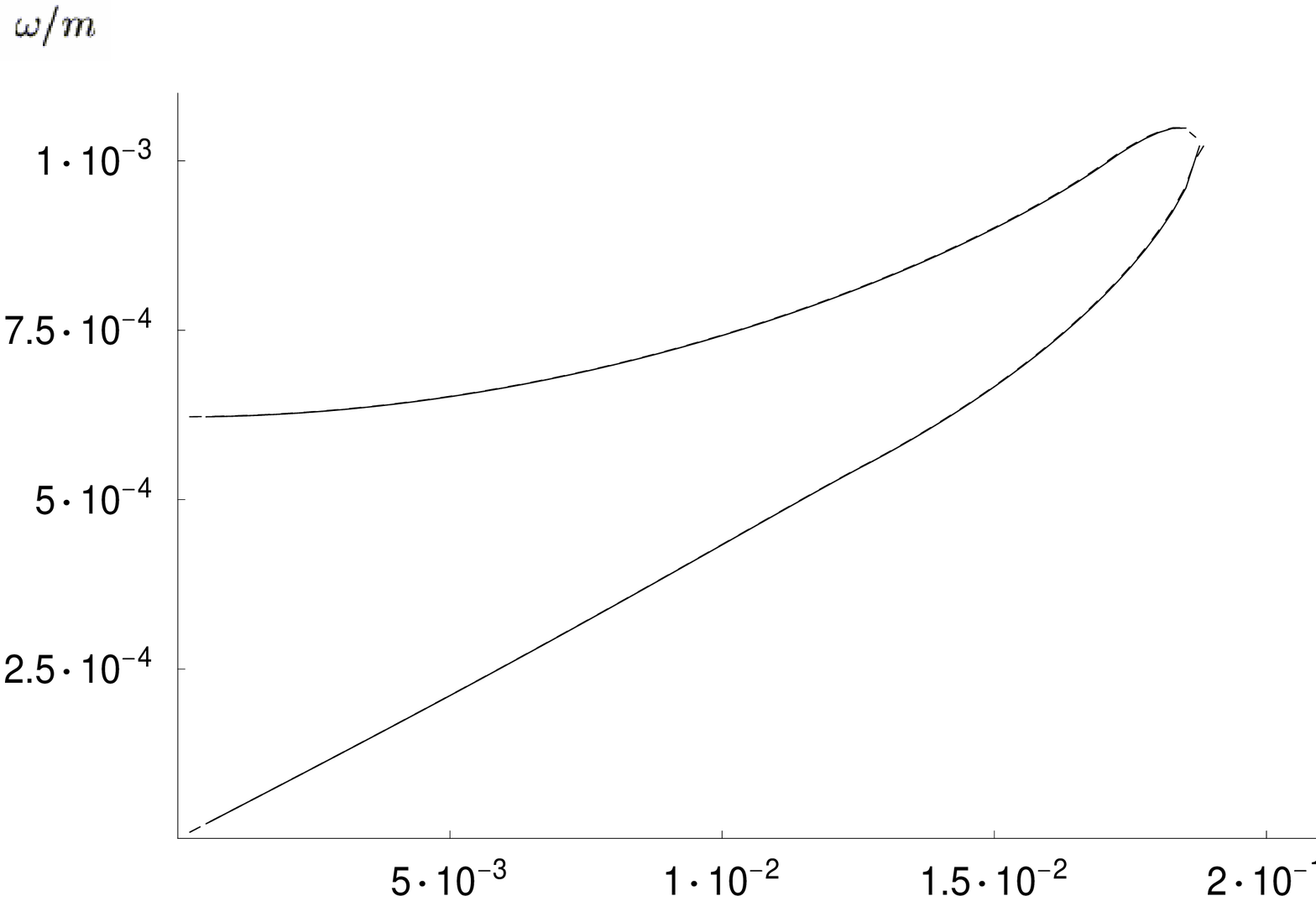}
\includegraphics[width=4.5in]{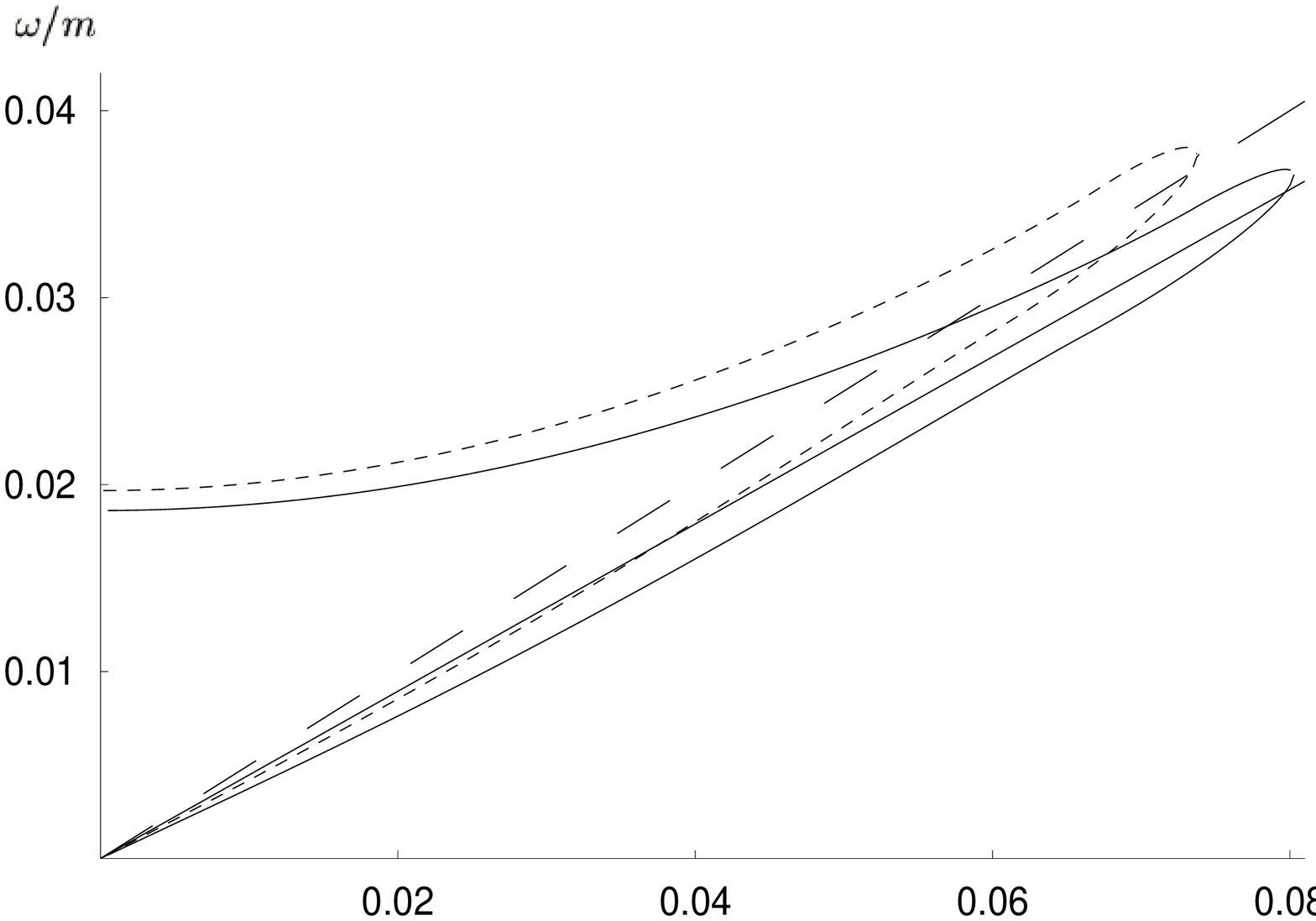}
\includegraphics[width=4.5in]{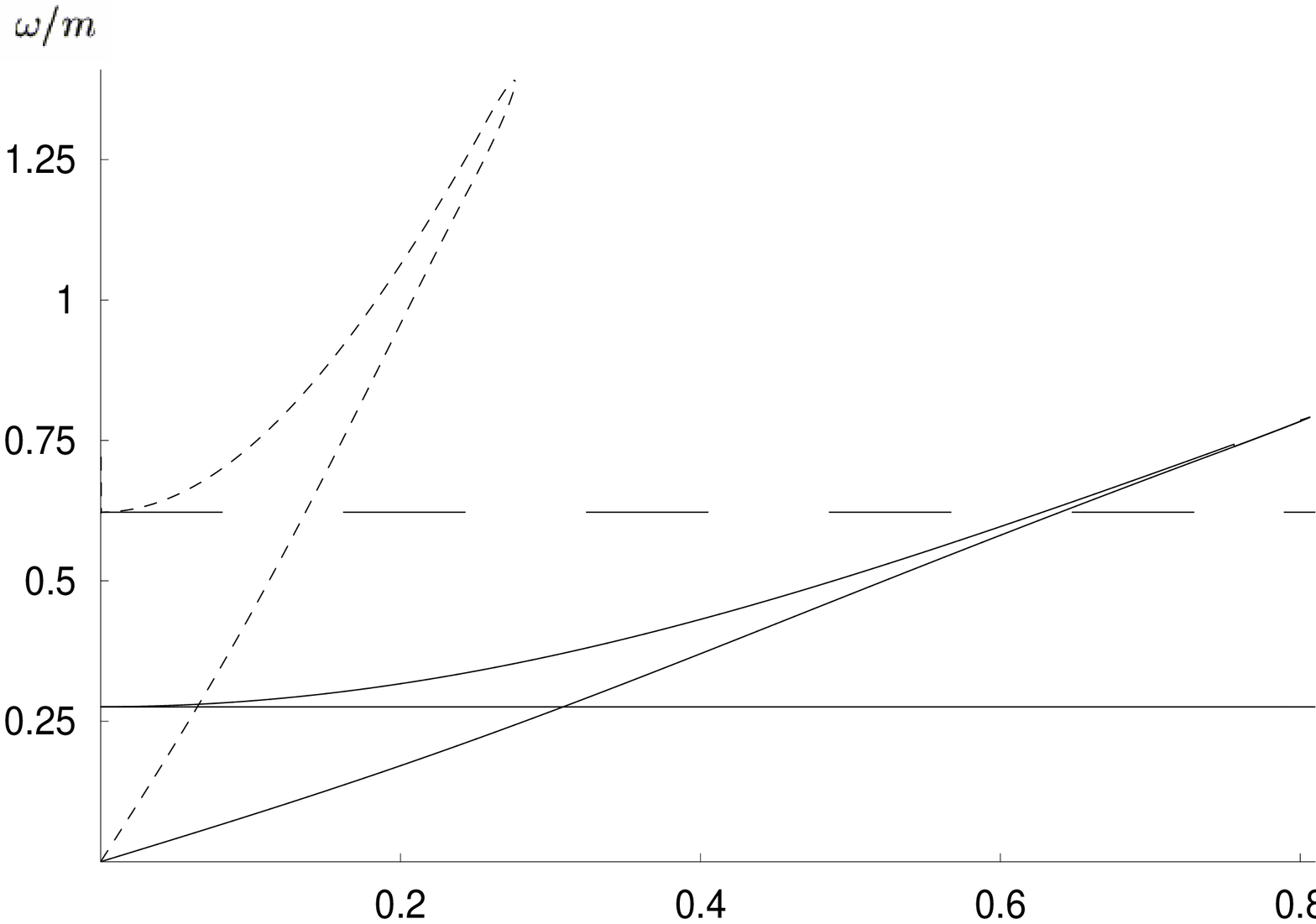}
\caption{The longitudinal dispersion relation for the Jancovici (solid) form and Lindhard (small dashed) form for increasing values of $\pf/m$. (a) For $\pf/m = 0.05$, the Jancovici and Lindhard forms are indistinguishable. (b)  For $\pf/m = 0.5$, the Jancovici and Lindhard forms are aligned along $\omega = |{\bf k}| \vf$ with $\vf=\pf/\ve_\rmF$ (solid straight line) and $\vf=\pf/m$ (long dashed line), respectively. (c) For $\pf/m = 5$, the cutoff frequency for the Jancovici form (solid curve) is obviously much lower than for the Lindhard form (long dashed curve); these are given by $\omega_c$, cf.\ (\ref{cutoff}), and $\omega_p$, respectively. }
\label{fig:long_modes_1}
\end{figure}

Numerical results are shown in Figure \ref{fig:long_modes_1} for $\pf/m = 0.05, 0.5, 5$. The wave modes are double valued, giving a tongue-like appearance of the dispersion curve, with the maximum frequency near the tip of the tongue, in agreement with \cite{KV91}. The dispersion curves calculated using the Lindhard and Jancovici response functions have negligible differences in the nonrelativistic regime, as shown for $\pf/m=0.05$. There are two obvious differences for larger $\pf$. One difference is that the tongue-like feature has a steeper slope with Lindhard's form than with Jancovici's form. This may be attributed to the tongue-like feature being aligned approximately along $\omega=|{\bf k}|v_\rmF$, with the Fermi speed given by $v_\rmF=p_\rmF/m$ in  Lindhard's form and by the relativistically correct value $v_\rmF=p_\rmF/\ve_\rmF$ in Jancovici's form. The other obvious difference is in the cutoff frequency. For Lindhard's form, the cutoff frequency is the plasma frequency, $\omega_p$, given by $\omega_p^2=e^2n_e/\varepsilon_0m$, with $n_e=\pf^3/3\pi^2$. With Jancovici's form, the cutoff frequency, $\omega_c$, is determined by \cite{HM84}
\be
\omega_c^2={4e^2m\over3\varepsilon_0\pi^2}
\int_0^{\pf}d|{\bi p}|\,|{\bi p}|^2\,
{3\varepsilon^2-|{\bi p}|^2\over4\varepsilon(\varepsilon^2-\omega_c^2)}.
\label{cutoff}
\ee
The cutoff frequency (\ref{cutoff}) reduces to the nonrelativistic plasma frequency in the nonrelativistic limit but, as in any relativistic electron gas,  is lowered by relativistic effects in a relativistic plasma. Note that some authors refer to the cutoff frequency as the `plasma frequency', but we adopt the more conventional definition in plasma physics, in which a distinction is made between the plasma frequency, $\omega_p$ as defined above, the proper plasma frequency, $\omega_{p0}$, defined in terms of the proper number density, $n_{e0}=[\vef\pf/m^2-\ln(\vef/m+\pf/m)]/2\pi^2$ here, and the cutoff frequency, determined by (\ref{cutoff}). The term $-\omega_c^2$ in the denominator on the right hand side of (\ref{cutoff}) is due to the relativistic quantum recoil; this term is neglected in all discussions of which we are aware. However, this neglect is well justified only for $\omega_c\ll m$.

\subsection{Transverse modes}
\begin{figure}[hp]
\includegraphics[width=4.5in]{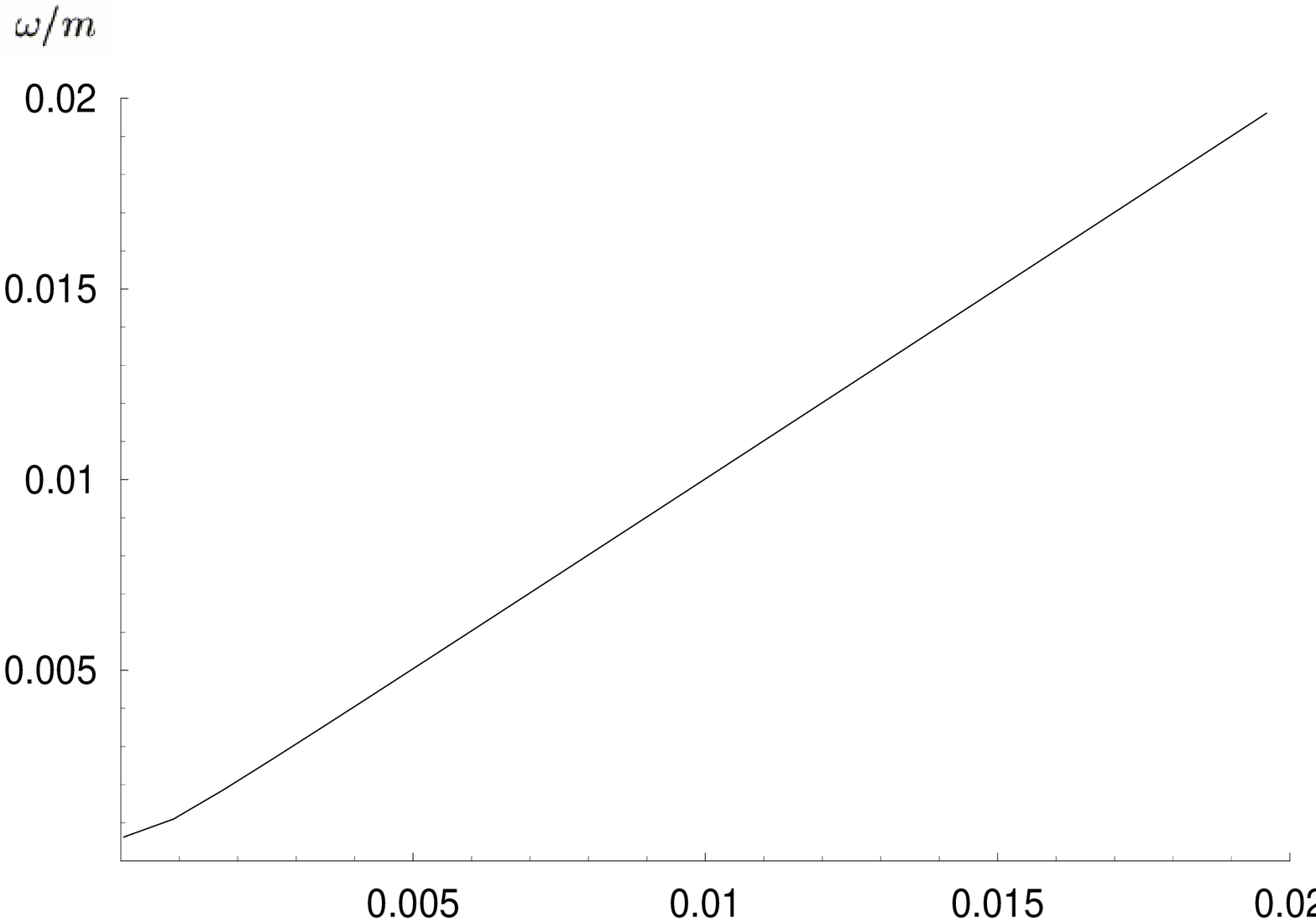}
\includegraphics[width=4.5in]{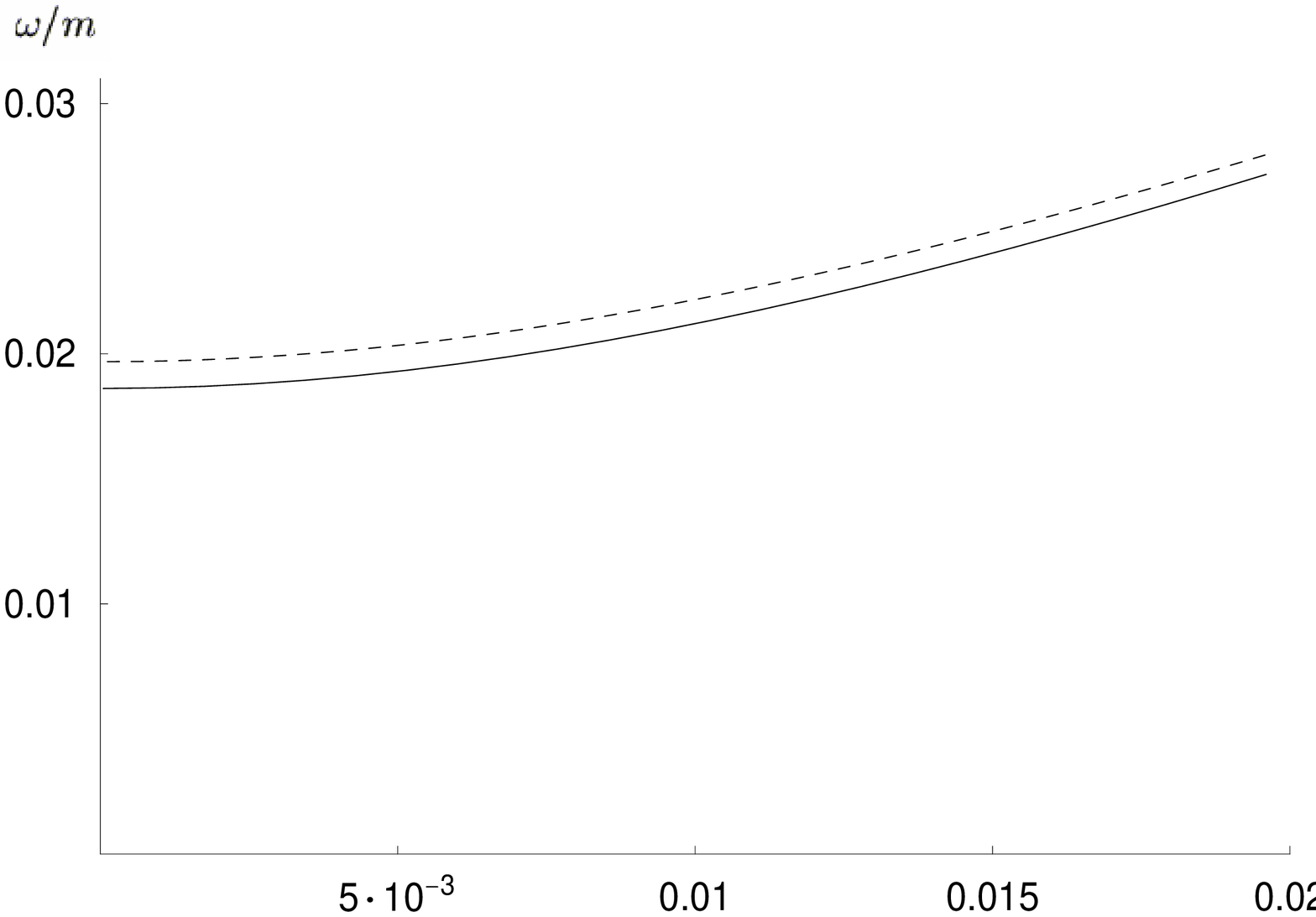}
\includegraphics[width=4.5in]{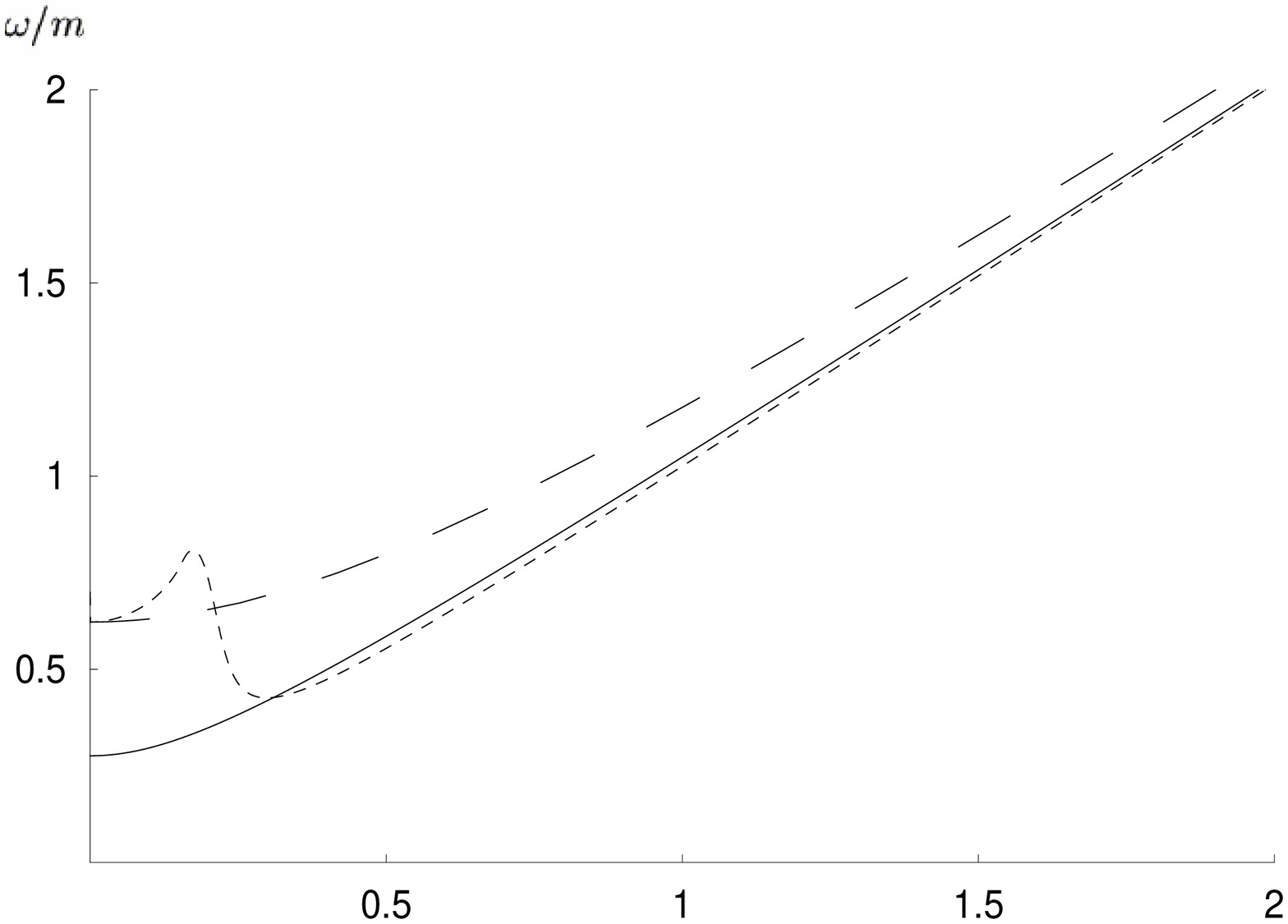}
\caption{The dispersion relations for the transverse mode for the Jancovici (solid) and Lindhard (small dashed) forms for increasing values of $\pf/m$. (a) For $\pf/m = 0.05$: the Jancovici and Lindhard are indistinguishable. (b) For $\pf/m = 0.5$, relativistic lowering of the dispersion curve is apparent, but the shape is essentially unchanged. (c) For $\pf / m =5$, with the cold plasma dispersion curve (long dashed) included for comparison: the Lindhard form jumps from the cold-plasma form to close to the Jancovici form for small $|{\bf k}|$. In all cases, the dispersion curve asymptotes to the light line at large $\bkm$.}
\label{fig:trans_modes_1}
\end{figure}

The dispersion relations for transverse modes are found by inserting the transverse response function  into (\ref{deqn1}), with the Jancovici and Lindhard forms given by (\ref{JT}) and (\ref{LT}), respectively. The numerical results are shown in Figure \ref{fig:trans_modes_1} for $\pf/m=0.05,0.5,5$. In figures \ref{fig:trans_modes_1}(a),  for $\pf/m=0.05$ the Jancovici and Lindhard forms are indistinguishable, and in (b) for $\pf/m=0.5$, relativistic effects cause a shift to lower frequencies compared with the nonrelativistic form. In figure \ref{fig:trans_modes_1}(c)  for $\pf/m=5$, a peculiar feature is apparent in the Lindhard form at small $|{\bf k}|$; the Lindhard form is close to the Jancovici form at higher frequencies but jumps rather abruptly to a cutoff at the plasma frequency due to this peculiar feature. The dispersion curve, $\omega^2 = \omega_p^2 + \bkm^2$, for a cold plasma is included for comparison.

\section{Discussion}

Our numerical results extend the known dispersion relation for longitudinal waves in a degenerate electron gas into the relativistic regime, and allow us to compare exact and approximate expressions for the dispersion relation for transverse waves in a completely degenerate relativistic electron gas. We comment here on the significance of these results.

\subsection{Dispersion curves}

The dispersion curve for longitudinal waves is known to have a tongue-like feature oriented along $\omega\approx\bkm\vf$ \cite{KV91}, where the Fermi speed, $\vf$, is related to the Fermi momentum by $\vf=\pf/m$ in the nonrelativistic case. We plot the generalization to the relativistic case, which also gives a tongue-like feature oriented along $\omega\approx\bkm\vf$, but with $\vf=\pf/\ve_\rmF$, with $\ve_\rmF=(m^2+\pf^2)^{1/2}$ the Fermi energy. The main changes from the nonrelativistic case are (a) this relativistically correct interpretation of $\vf$, (b) the decrease of the cutoff frequency, and (c) a sharpening of the peak in the dispersion curve where it turns over. These features are illustrated in Figure~\ref{fig:long_modes_1} where exact dispersion curves are compared with their nonrelativistic counterparts.

The form of the dispersion curve for transverse waves in any isotropic plasma is severely constrained by the requirements that the frequency be equal to the cutoff frequency for $\bkm=0$ and asymptote to the light line, $\omega=\bkm$, for $\bkm\to\infty$. We compare the dispersion relations for transverse waves derived using Jancovici's and Lindhard's transverse response functions, and find there to be an apparently spurious feature in the latter. A test on the validity of the transverse response function is that it gives the correct result for the magnetic susceptibility of an electron gas, which is known to have the same ratio of contributions from Pauli spin paramagnetism and Landau orbital diamagnetism in the relativistic case as in the familiar nonrelativistic case \cite{Silin}. Jancovici's transverse response function reproduces this results but Lindhard's does not. This failure may be attributed to the somewhat artificial procedure for including the spin of the electron in Lindhard's calculation; this was recognized by Lindhard himself in that his transverse response function fails to reproduce the correct value for the magnetic susceptibility of an electron gas. This peculiar feature in the dispersion curve for the Lindhard form is non-physical, and is due to an incorrect treatment of the electron spin.

\subsection{PC for transverse waves in superdense plasmas}

The properties of waves, especially transverse waves, in a superdense plasma has led to some controversy. Our results help clarify one aspect of this controversy, but there are two other aspects of it that we have not resolved.

\begin{figure}[ph]
\centering\includegraphics[width=5in]{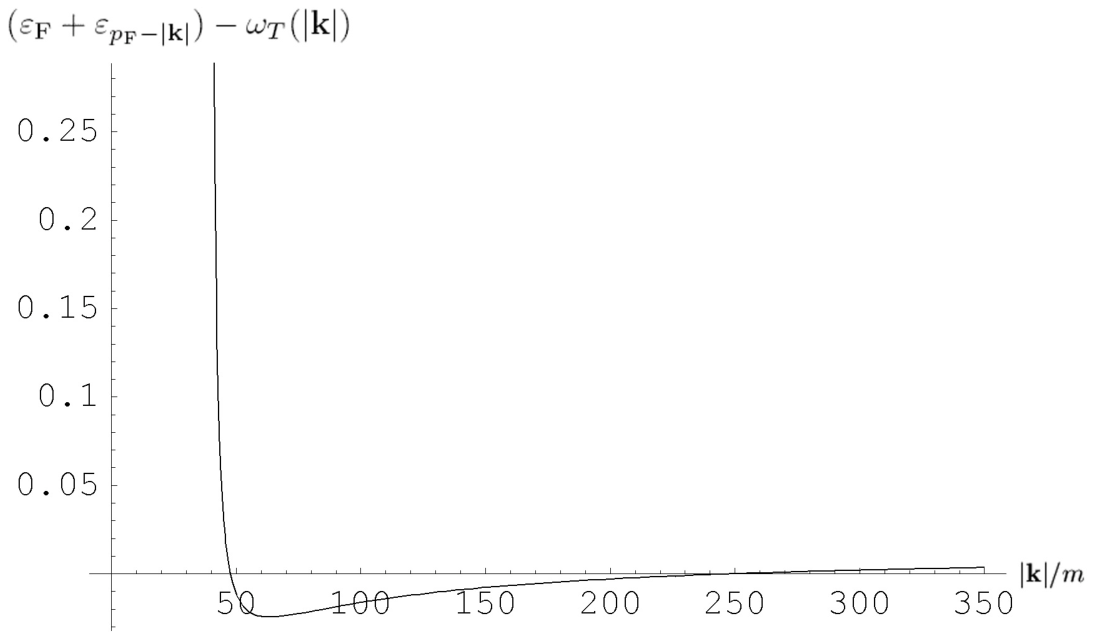}
\caption{A plot showing the difference, $\vef - \ve_{\pf-\bkm} -\omega_T(\bfk)$, between the boundary curve of the region where PC is allowed and the dispersion curve for the Jancovici transverse mode for $\pf/m = 40$. The cutoff frequency is above the threshold, $2m$, for PC in vacuo, but in a region where PC in the vacuum is completely suppressed by the electron. At $\bkm/m = 47.7$ the difference becomes negative and the dispersion curve mode is in a region where PC is possible. At $\bkm/m =247.9$ the dispersion curve enters a region where PC has its full vacuum value. }
\label{fig:janco_trans_mode_regime}\end{figure}

A controversial point is whether PC is possible for waves in a superdense plasma. Early authors \citep{T61,HM84} assumed that the cutoff frequency could exceed the threshold for PC, $\omega_c>2m$, and that there is then a portion of the dispersion curve just above the cutoff frequency in the region where PC is allowed. These discussions were for an arbitrary electrons gas, and were not restricted to the completely degenerate case. In Ref.~\cite{neutrino0} it was pointed out that PC has important implications for the plasma process for neutrino emission from dense plasmas.  However, the claim that PC is allowed was declared to be untrue in Ref.~\cite{B91}.  In Ref.\ \cite{IMHK92} it was pointed out that for a completely degenerate electron gas, the PC threshold is $\vef+m$, rather than $2m$, because the only available electron states are above $\vef$. It was further argued that the cutoff frequency (which was approximated by the proper plasma frequency in Ref.~\cite{IMHK92}) can never exceed the threshold frequency for PC. Our results imply existence of a DF region in Figure~\ref{fig:thresholds} in the range $2m<\omega<m+\vef$, which confirms the point made in Ref.\ \cite{IMHK92}, but only for $\bkm=0$. As is evident from Figure~\ref{fig:thresholds}, and as we point out elsewhere \cite{MWM05}, this region shrinks with increasing $\bkm$ and is absent for $\bkm>2\pf$. We conclude that in superdense plasmas, defined to satisfy $\omega_c>2m$, although the dispersion curve has its cutoff in the DF region  $2m<\omega<m+\vef$, the dispersion curve enters the region where PC is allowed at a higher $\bkm$. Hence, at sufficiently high $\omega$, a photon (in a completely degenerate electron gas) can decay into a pair and a pair can annihilate into a single photon. In Figure~\ref{fig:janco_trans_mode_regime} we plot the difference between the dispersion curve and the boundary of the DF region to illustrate this effect for a particular value of $\pf$.

Our results lead us to conclude that PC is possible in a superdense plasma, but this conclusion is subject to two important provisos. The first proviso is that we neglect mass renormalization. The modification of the mass of the electron was central to the original argument in  Ref.\ \cite{B91}, but has been ignored in most subsequent discussions. Moreover, the form of the mass renormalization needs further investigation: in existing treatments the photon propagator is taken to be that in vacuo, whereas the photon propagator in the plasma should be used. The other proviso concerns the evaluation of the cutoff frequency. The cutoff frequency is determined by (\ref{cutoff}) and in all existing discussions, the term $\omega_c^2$ in the denominator of the integrand is neglected. This is only justified for $\omega_c^2\ll m^2$, and this is not necessarily satisfied in a superdense plasma. For example, in Ref.\ \cite{BS93} the semi-classical forms (\ref{BSL}) and (\ref{BST}) were used to treat the dispersion, and while this is valid for transverse waves well above the cutoff frequency, it is not valid in a superdense plasma near the cutoff frequency, because the $-\omega_c^2$ on the right hand side of (\ref{cutoff}) is due to the quantum recoil term that is neglected in the semi-classical approximation.

In a more general discussion of wave dispersion in a superdense plasma one needs to include both the mass renormalization and the (relativistically correct) quantum recoil in a self-consistent manner. This involves a major generalization of existing theory that we do not attempt here. The question as to whether or not PC is allowed in a superdense plasma should remain open until such a general treatment is available.

\subsection{Pair modes}

Inclusion of an additional source of dispersion in a plasma usually leads to the appearance of intrinsically new wave modes associated with that dispersion, and pair modes are associated with dispersion due to PC. Pair modes are known to exist in degenerate Bose gases \cite{KFH85,WM89}, and in a magnetized completely degenerate electron gas \cite{PK92}. Although we searched for pair mode solutions we found none in an isotropic degenerate electron gas. In a degenerate Bose gas, the Bose condensate contributes a pole in the dispersion function, and the existence of pair modes results directly from the presence of the pole. In a magnetized completely degenerate electron gas, there is a logarithmic singularity in the response functions associated with the threshold for PC, and the existence of pair modes is attributed to this feature \cite{PK92}. Our plots of the response functions for an unmagnetized completely degenerate electron gas show no singular feature associated with the threshold for PC, cf.~Figures~\ref{fig:janco_long_log} and~\ref{fig:janco_trans_log}. Our failure to find solutions corresponding to pair modes is consistent with the suggestion in Ref.~\cite{PK92} that they are associated with singularities in (the real parts of) the response functions at the threshold for PC. 

\section{Conclusions}

We compare the response functions for a completely degenerate electron gas in three forms: the fully relativistic form due to Jancovici \cite{J62}, the nonrelativistic form due to Lindhard \cite{L54}, and the semi-classical approximation in which the quantum recoil is neglected. An important physical difference between the Jancovici and Lindhard (and semi-classical) forms is the existence in the relativistic case of one-photon pair creation (PC) as a dissipation mechanism, in addition to Landau damping (LD). This leads to a PC-associated dispersion that is intrinsic to the relativistic quantum case. We plot the response functions and find that they vary smoothly as the thresholds for LD and PC are crossed. The semi-classical approximation, cf.\ (\ref{BSL}) and (\ref{BST}), is found to be accurate for $\bkm\ll2\pf$, but not for $\bkm\gg2\pf$ (except for $\omega/\bkm\to1$).

We plot dispersion curves for longitudinal and transverse waves. The relativistic treatment leads to a natural generalization of a known result in the nonrelativistic case \cite{KV91}: the dispersion curve for longitudinal waves has a tongue-like appearance, with the Langmuir-like branch roughly along $\omega=\bkm\vf$, turning over at a maximum frequency and joining onto the zero-sound branch. The main relativistic effects are that the Fermi speed is replaced by its relativistically correct value, $\vf=\pf/\ve_\rmF$, and the cutoff frequency by its relativistically correct value (\ref{cutoff}). Both the longitudinal and transverse modes cutoff at the same frequency, which is close to but not equal to the proper plasma frequency, with the proviso that this is well below the threshold for PC. For transverse waves, comparison of the Jancovici and Lindhard forms shows that they are similar except for a nonphysical feature in the Lindhard case near the cutoff frequency for high $\pf$, where the dispersion curve jumps from near the relativistically correct cutoff to the nonrelativistic cutoff. This feature appears to be associated with an intrinsic weakness in the Lindhard form, which does not treat the spin correctly and does not reproduce the well-known form for the magnetic susceptibility.

We comment on a controversy concerning whether or not PC is possible in a superdense plasma, where the plasma frequency exceeds the threshold for PC. Contrary to earlier claims \cite{T61,neutrino0,HM84} that PC is possible, it was argued in Refs~\cite{B91,IMHK92,BS93} that it is not possible. Our results show that the dispersion curves do enter the region where PC is allowed, and we point out deficiencies in the arguments to the contrary. However, the conclusion that PC is possible in a superdense plasma is subject to two important provisos: mass renormalization is neglected, and the cutoff frequency is approximated by neglecting the term $\omega_c2$ in the denominator on the right hand side of (\ref{cutoff}). In any more detailed investigation of dispersion in a superdense plasma, both these effects should be taken into account simultaneously.

We searched for pair modes, known to exist in degenerate Bose gases \cite{KFH85,WM89} and a degenerate magnetized electron gas \cite{PK92}. We found no evidence for pair modes in an isotropic degenerate electron gas. Pair modes in a degenerate Bose gas are associated with the contribution to the response from the Bose condensate, which gives a pole in the response functions; pair modes in a magnetized degenerate electron gas are associated with a logarithmic singularity at the threshold for PC \cite{PK92}. There is no such singularity in the response functions for an isotropic electron gas. 

\section*{\it Acknowledgments}
\noindent We thank Sergey Vladimirov for helpful discussions and Qinghuan Luo for helpful comments on the manuscript.

\end{document}